\documentclass[12pt]{article}
\usepackage{amsmath}
\usepackage{psfrag}
\usepackage{graphicx}
\usepackage{graphics}
\usepackage{epsfig}
\newcommand{\as}{a\!\!\!/}
\newcommand{\As}{A\!\!\!/}
\newcommand{\ks}{k\!\!\!/}
\newcommand{\ps}{p\!\!\!/}

\date{}
\begin{document}
\baselineskip=18.6pt plus 0.2pt minus 0.1pt \makeatletter
%\@addtoreset{equation}{section} \renewcommand{\theequation}{\thesection.%
%\arabic{equation}} \makeatletter \@addtoreset{equation}{section}

%\begin{titlepage}

\title{\vspace{-3cm}
\hfill\parbox{4cm}{\normalsize \emph{}}\\
 \vspace{1cm}
{Electron's anomalous magnetic moment effects on laser assisted ionization of atomic hydrogen by electronic impact}}
 \vspace{2cm}
\author{S. Taj$^1$, B. Manaut$^1$\thanks{{\tt b.manaut@usms.ma}}, M. El Idrissi$^1$, Y. Attaourti$^2$ and L. Oufni$^3$\\
{\it {\small $^1$ Facult\'e Polydisciplinaire, Universit\'e Sultan Moulay Slimane,}}\\
{\it {\small Laboratoire Interdisciplinaire de Recherche en Sciences et Techniques (LIRST),}}\\
 {\it {\small Boite Postale 523, 23000 B\'eni Mellal, Morocco.}}\\
 {\it {\small $^2$ Universit\'e Cadi Ayyad, Facult\'e des Sciences Semlalia,}}\\
  {\it {\small D\'epartement de Physique, LPHEA, BP : 2390, Marrakech,
  Morocco.}}\\
  {\it {\small $^3$ Universit\'e Sultan Moulay Slimane, Facult\'e des Sciences et Techniques,}}\\
  {\it {\small D\'epartement de Physique, LPMM-ERM, BP : 523, 23000 B\'eni Mellal,
  Morocco.}}}
   \maketitle \setcounter{page}{1}
\begin{abstract}
Electron-impact ionization of atomic hydrogen with the electron's
anomalous magnetic moment (AMM) effects are examined. The formulas
for the laser-assisted relativistic triple differential cross
section (TDCS) in the coplanar binary geometry developed earlier by
Y. Attaourti and S. Taj [Phys. Rev. A \textbf{69}, 063411 (2004)]
are used to check the consistency of our computations when the
anomaly $\kappa$ is taken to be zero. We show that the contribution
of the terms containing the AMM effects even in the first Born
approximation has an important contribution, so it must be included
in any reliable analysis. A full analytical calculation for the TDCS
is presented.\vspace{0.1in}

PACS number(s): 34.80.Qb, 34.50.Rk, 34.50.Fa, 12.20.Ds
\end{abstract}

\maketitle
\section{Introduction}
The hydrogen atom, due to its simplicity, has a central role in the understanding of chemistry and atomic physics. Apart from the fundamental interest, reactions involving atomic hydrogen have practical importance in controlled thermonuclear fusion and in the earth's high atmosphere. Ionization of atoms or ions in collision by charged particles is important for diagnostics of high temperature plasmas as well as for fundamental understanding of the atomic structure. In recent years, the electron coincidence spectroscopy
has become a powerful tool for testing dynamic theories of final states with
two outgoing electrons[1-4]. To the author's knowledge, no relativistic experimental data of the electron-impact ionization of atoms in laser assisted has been given. A complete kinematic analysis can provide an overall symmetry of the impact ionization
processes and facilitate the comparison between theory and experiment.
The theoretical models developed through the years to uncover the details of these
processes can be classified according to the impact energy of the projectile. In the low-to-
intermediate energy regime, we found close-coupling methods based on a molecular approach
[5,6]; in the intermediate regime, the classical trajectory Monte Carlo has been widely used [7],
while in the intermediate-to-high energy regime, the distorted wave method can be applied
[8,9]. Furthermore, there are methods that can be applied throughout the complete range of
impact energies [10-12]. Though the distorted wave method offers several advantages like including
the correct asymptotic conditions of the wave functions due to the long-range behavior of
the Coulomb interaction between the particles [13]; it doesn't take into account the electron's anomalous magnetic moment effects. In this paper, we present a theoretical model for the relativistic electronic dressing in laser-assisted ionization of atomic hydrogen by electron impact with electron's anomalous magnetic moment effects. For pedagogical purposes, in section 2 we begin our study without AMM effects (electron's anomaly is taken to be zero). In section 3, we present our study with AMM effects. In section 4, we discuss the results we have obtained. Throughout this work, we use atomic units $\hbar=m=e=1$ and work with the metric
tensor $g^{\mu\nu}=g_{\mu\nu}=diag(1,-1,-1,-1)$. In many equations
of this paper, the Feynman 'slash notation' is used. For any
4$-$vector $A$, $\As=A^{\mu}\gamma_{\mu}=A^{0}\gamma_{0}-\mathbf{A}.\mathbf{\gamma}$
where the matrices $\gamma$ are the well known Dirac matrices.
\section{THE TDCS IN LASER ASSISTED WITHOUT AMM.}
We now take into account the electronic relativistic dressing
of all electrons which are described by Dirac-Volkov
plane waves normalized to the volume $V$. This gives rise to a trace already given in detail in \cite{14} but it will turn out that taking into account the relativistic electronic dressing of the ejected electron amounts simply to introduce a new sum on the $l_B$
photons that can be exchanged with the laser field. The transition amplitude is now given by
{\small \begin{eqnarray}
S_{fi}=-\frac{i}{c}\int_{-\infty }^{+\infty }dx^{0}\langle\psi
_{q_{f}}(x_{1})\phi _{f}(x_{2})\mid V_{d}\mid \psi
_{q_{i}}(x_{1})\phi _{i}(x_{2})\rangle. \label{1}
\end{eqnarray}}
where the Dirac-Volkov wave function for the
ejected electron reads as
\begin{equation}
\phi _{f}(x_{2})=\psi _{q_{B}}(x_{2})=[1+\frac{\ks\As_{(2)}}{2c(k.p_{B})}]\frac{%
u(p_{B},s_{B})}{\sqrt{2Q_{B}V}}e^{is_{B}(x_{2})},\label{2}
\end{equation}
where $A_{(2)}=a_{1}\cos (\phi _{2})+a_{2}\sin (\phi _{2})$ is
the four potential of the laser field felt by the ejected
electron, $\phi
_{2}=k.x_{2}=k_{0}x_{2}^{0}-\mathbf{k.x}_{2}=wt-\mathbf{k.x}_{2}$
is the phase of the laser field and $w$ its pulsation. \\
Proceeding along the same line as before, we obtain for the spin-unpolarized triple differential cross
section evaluated for $Q_{f}=Q_{i}+(s+l_{B})w+\varepsilon _{b}-Q_{B}$ the following formula
{\small
\begin{eqnarray}
\frac{d\overline{\sigma
}}{dE_{B}d\Omega_{B}d\Omega_{f}}=\sum_{s,l_{B}=-\infty }^{+\infty}\frac{d\overline{\sigma
}^{(s,l_{B})}}{dE_{B}d\Omega _{B}d\Omega _{f}},\label{3}
\end{eqnarray}}
with
{\small
\begin{eqnarray}
\frac{d\overline{\sigma }^{(s,l_{B})}}{dE_{B}d\Omega _{B}d\Omega
_{f}} &=&
\frac{1}{2}\frac{|\mathbf{q}_{f}||\mathbf{q}_{B}|}{|\mathbf{q}_{i}|c^{6}}\frac{\left(\frac{1}{2}\sum_{s_{i},s_{f}}\mid
M_{fi}^{(s)}\mid ^{2}\right)}{\mid
\mathbf{q}_{f}\mathbf{-q}_{i}\mathbf{-}s \mathbf{k}\mid
^{4}}\sum_{s_{B}}\mid \overline{u}(p_{B},s_{B})\nonumber\\
&&\times\Gamma _{l_{B}}\gamma^{0}\mid ^{2}
\mid\Phi _{1,1/2,1/2}(\mathbf{q=\Delta
}_{s+l_{B}}\mathbf{-q}_{B})\nonumber\\
&&-\Phi
_{1,1/2,1/2}(\mathbf{q=-q}_{B}+l_{B}\mathbf{k})\mid ^{2}.\label{4}
\end{eqnarray}
}
The sum $ ( \sum_{s_{i},s_{f}}\mid
M_{fi}^{(s)}\mid ^{2}/2) $ has already been evaluated in a
previous work \cite{15}. The quantity $\mathbf{\Delta }_{s+l_{B}}$ is simply given by $\mathbf{\Delta }
_{s+l_{B}}=\mathbf{q}_{i}\mathbf{-q}_{f}+(s+l_{B})\mathbf{k}$.
Introducing the factor $c(p_{B})=1/(2c(k.p_{B}))$, the symbol
$\Gamma _{l_{B}}$ is defined as
\begin{equation}
\Gamma _{l_{B}}=B_{l_{B}}(z_{B})+c(p_{B})[\as_{1}\ks
B_{1l_{B}}(z_{B})+\as_{2}\ks B_{2l_{B}}(z_{B})],\label{5}
\end{equation}
where the three quantities $B_{l_{B}}(z_{B})$, $B_{1l_{B}}(z_{B})$ and $
B_{2l_{B}}(z_{B})$ are respectively given by
{\small
\begin{eqnarray}
\begin{array}{c}
  B_{l_{B}}(z_{B})=J_{l_{B}}(z_{B})e^{il_{B}\phi _{0B}} \\
  B_{1l_{B}}(z_{B})=\{J_{l_{B}+1}(z_{B})e^{i(l_{B}+1)\phi
_{0B}}+J_{l_{B}-1}(z_{B})e^{i(l_{B}-1)\phi _{0B}}\}/2\\
  B_{2l_{B}}(z_{B})=\{J_{l_{B}+1}(z_{B})e^{i(l_{B}+1)\phi
_{0B}}-J_{l_{B}-1}(z_{B})e^{i(l_{B}-1)\phi
_{0B}}\}/2i ,
\end{array} \label{6}
\end{eqnarray}}
where $z_{B}=\frac{|a|}{ c(k.p_{B})}\sqrt{(%
\widehat{\mathbf{y}}\mathbf{.p}_{B})^{2}+(\widehat{\mathbf{x}}\mathbf{.p}
_{B})^{2}}$
 is the argument of the ordinary Bessel functions
that will appear in the calculations and the phase $\phi _{0B}$
is defined by
\begin{equation}
\phi _{0B}=\arctan ((\widehat{\mathbf{y}}\mathbf{.p}_{B})/(\widehat{\mathbf{x%
}}\mathbf{.p}_{B})).
\end{equation}
The sum over the spins of the ejected electron can be transformed
to traces of gamma matrices. Using REDUCE \cite{16}, we find
\begin{eqnarray}
&&\sum_{s_{B}} \mid \overline{u}(p_{B},s_{B})\Gamma
_{l_{B}}\gamma ^{0}\mid
^{2}=4\{E_{B}J_{l_{B}}^{2}(z_{B}) \nonumber\\
&&+wc(p_{B})(\cos (\phi _{0B})(a_{1}.p_{B})+\sin (\phi
_{0B})(a_{2}.p_{B}))
\nonumber\\
&&\times J_{l_{B}}(z_{B})(J_{l_{B}+1}(z_{B})+J_{l_{B}-1}(z_{B}))\nonumber \\
&&-a^{2}w(k.p_{B})c^{2}(p_{B})(J_{l_{B}+1}^{2}(z_{B})+J_{l_{B}-1}^{2}(z_{B}))\}.\label{7}
\end{eqnarray}
As expected, in the absence of the laser field only the term $4E_{B}J_{l_{B}}^{2}(z_{B}=0)%
\delta _{l_{B},0}=4E_{B}$ contributes to the TDCS.
\section{THE TDCS IN LASER ASSISTED WITH AMM}
 We now take into account the AMM effects of all electrons (incident, scattered and ejected) which are described in the weak-field approximation (WFA) \cite{17} by :
{\small
\begin{eqnarray}
\psi(x)=&&[1-(\alpha \ks\As+\beta\ks+\delta\ps\ks\As)]\frac{u(p,s)}{\sqrt{2VQ_0}}\nonumber\\
&&\times\exp\left[-i(qx)-i\int_0^{kx}\frac{(Ap)}{c(kp)}d\phi\right]\label{8}
\end{eqnarray}}
with
\begin{eqnarray}
\alpha=\frac{1}{2(k.p)}\left(\frac{\kappa_l c}{2}-\frac{1}{c}\right)\; ;\;\beta=\frac{\kappa_l A^2}{4c(k.p)}\; ;\; \delta=\frac{\kappa_l}{4(k.p)}\label{9}
\end{eqnarray}
The transition amplitude with AMM effects is given by
{\small \begin{eqnarray}
S^{AMM}_{fi}=-\frac{i}{c}\int_{-\infty }^{+\infty }dx^{0}\langle\psi
_{q_{f}}(x_{1})\phi _{f}(x_{2})\mid V_{d}\mid \psi
_{q_{i}}(x_{1})\phi _{i}(x_{2})\rangle.\label{10}
\end{eqnarray}}
The spin-unpolarized triple differential cross section with the AMM effects evaluated for $Q_{f}=Q_{i}+(s+l_{B})w+\varepsilon _{b}-Q_{B}$ is given by :
{\small
\begin{eqnarray}
\frac{d\overline{\sigma
}^{AMM}}{dE_{B}d\Omega_{B}d\Omega_{f}}&=&
\frac{1}{2}\frac{|\mathbf{q}_{f}||\mathbf{q}_{B}|}{|\mathbf{q}_{i}|c^{6}}\frac{\left(\frac{1}{2}\sum_{s_{i},s_{f}}\mid
M_{fi}^{(s)}\mid ^{2}\right)}{\mid
\mathbf{q}_{f}\mathbf{-q}_{i}\mathbf{-}s \mathbf{k}\mid
^{4}}\sum_{s_{B}}\mid \overline{u}(p_{B},s_{B})\nonumber\\
&&\times\Delta _{l_{B}}\gamma^{0}\mid ^{2}
\mid\Phi _{1,1/2,1/2}(\mathbf{q=\Delta
}_{s+l_{B}}\mathbf{-q}_{B})\nonumber\\
&&-\Phi
_{1,1/2,1/2}(\mathbf{q=-q}_{B}+l_{B}\mathbf{k})\mid ^{2}.\label{11}
\end{eqnarray}
}
The spinorial part  $ ( \sum_{s_{i},s_{f}}\mid M_{fi}^{(s)}\mid ^{2}/2) $ is the factor in which electron's AMM effects are reflected \cite{18}. However, the novelty in the various stages of the calculations when including the AMM effects of the ejected electron is contained in  the symbol $\Delta_{l_{B}}$ which is given as
{\small\begin{eqnarray}
\Delta_{l_{B}}=&&(1-\ks\beta_B)B_{l_{B}}(z_{B})-[\delta_B\as_{1}\ks\ps_B+\alpha_B\as_{1}\ks]B_{1l_{B}}(z_{B})\nonumber\\
&&-[\delta_B\as_{2}\ks\ps_B+\alpha_B\as_{2}\ks]B_{2l_{B}}(z_{B})],\label{12}
\end{eqnarray}}
where the three quantities $B_{l_{B}}(z_{B})$, $B_{1l_{B}}(z_{B})$ and $
B_{2l_{B}}(z_{B})$ are given in Eq. (\ref{6}).\\
The sum over the spin of the ejected electron can be transformed
to traces of gamma matrices. Using REDUCE \cite{16} and after tedious calculations, we obtain the trace of the ejected electron in it's final form
{\small
\begin{eqnarray}
&&\sum_{s_{B}} \mid \overline{u}(p_{B},s_{B})\Delta
_{l_{B}}\gamma ^{0}\mid
^{2}=\frac{1}{2c^2 (k.p_B)}[-4 c^2\kappa a^2\omega+8 c^2 (k.p_B)\nonumber \\
&&\times  E_B+ \kappa^2 a^2\omega]J_{l_{B}}^{2}(z_{B})+\frac{\omega}{2 c (k.p_B)} [-\cos(\phi_{0B})(a_1.p_B) \kappa^2 a^2\nonumber\\
&&+4\cos(\phi_{0B}) (a_1.p_B)- \sin(\phi_{0B}) (a_2.p_B) \kappa^2 a^2+4\sin(\phi_{0B})\nonumber \\
&&\times(a_2.p_B)] J_{l_{B}}(z_{B})(J_{l_{B}+1}(z_{B})+J_{l_{B}-1}(z_{B}))+\frac{a^2}{2 c^2(k.p_B)}\nonumber \\
&&\times [-c^2 \kappa^2 (k.p_B) E_B+2 c^2\kappa \omega-2\omega](J_{l_{B}+1}^{2}(z_{B})+J_{l_{B}-1}^{2}(z_{B})).\nonumber \\
\end{eqnarray}}
The first check to be done is to take $\kappa=0$ in order to recover all the results in the absence of
the anomalous magnetic moment effect. When this is done, one recovers the simple trace's result given in Eq. (\ref{7}). Once again, when no radiation field is present, this trace reduces to $4E_{B}J_{l_{B}}^{2}(z_{B}=0)\delta _{l_{B},0}=4E_{B}$.
\begin{figure}[!h]
\centering
\includegraphics[angle=0,width=3in,height=2.5 in]{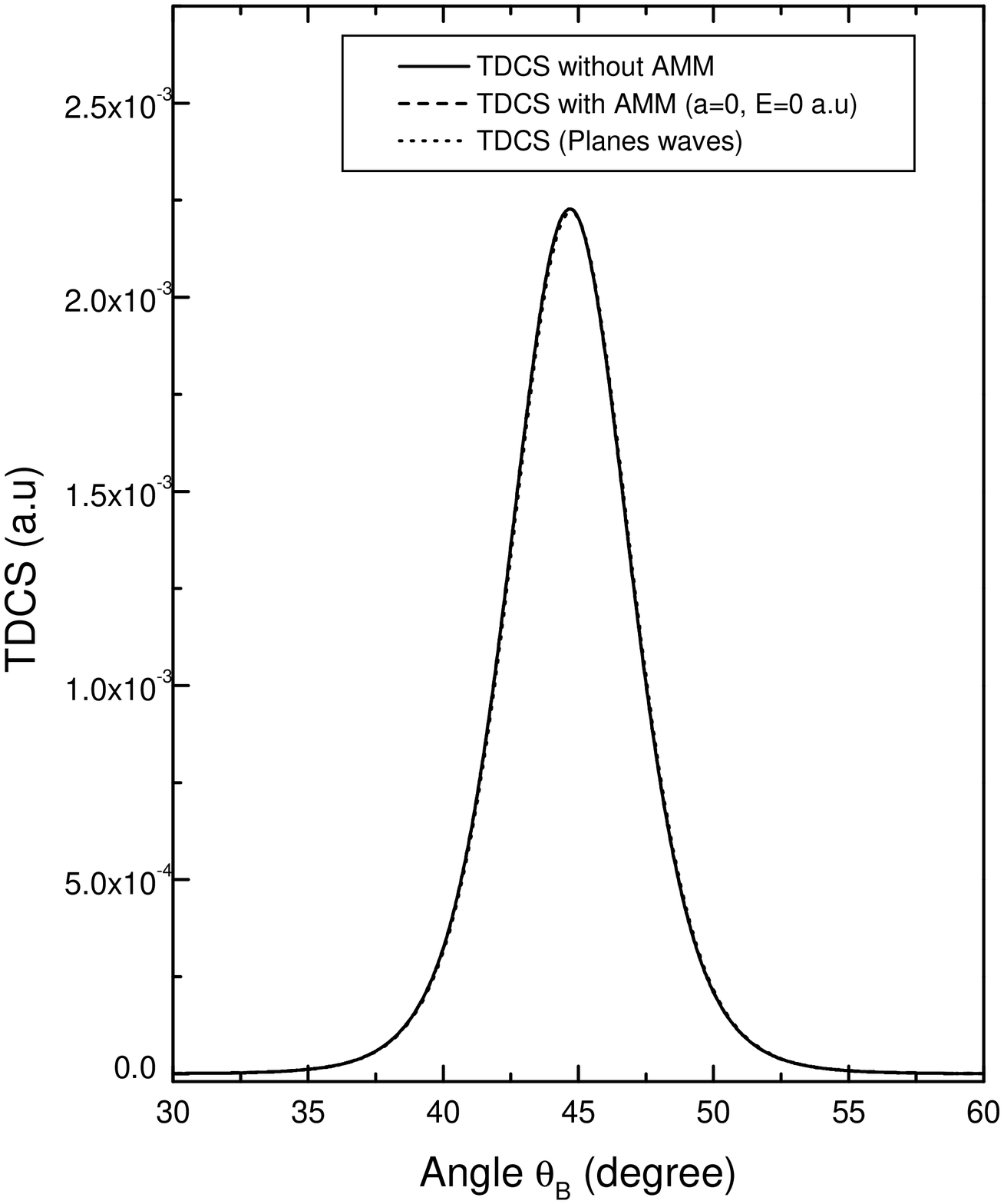}
\caption{The TDCSs as a function
of the angle $\theta_B$. The incident electron kinetic
energy is $T_i=2700 \;eV$ and the ejected electron kinetic energy is $T_B=1349.5 \;eV$.  }
\end{figure}
\begin{figure}[!h]
\centering
\includegraphics[angle=0,width=3in,height=2.5 in]{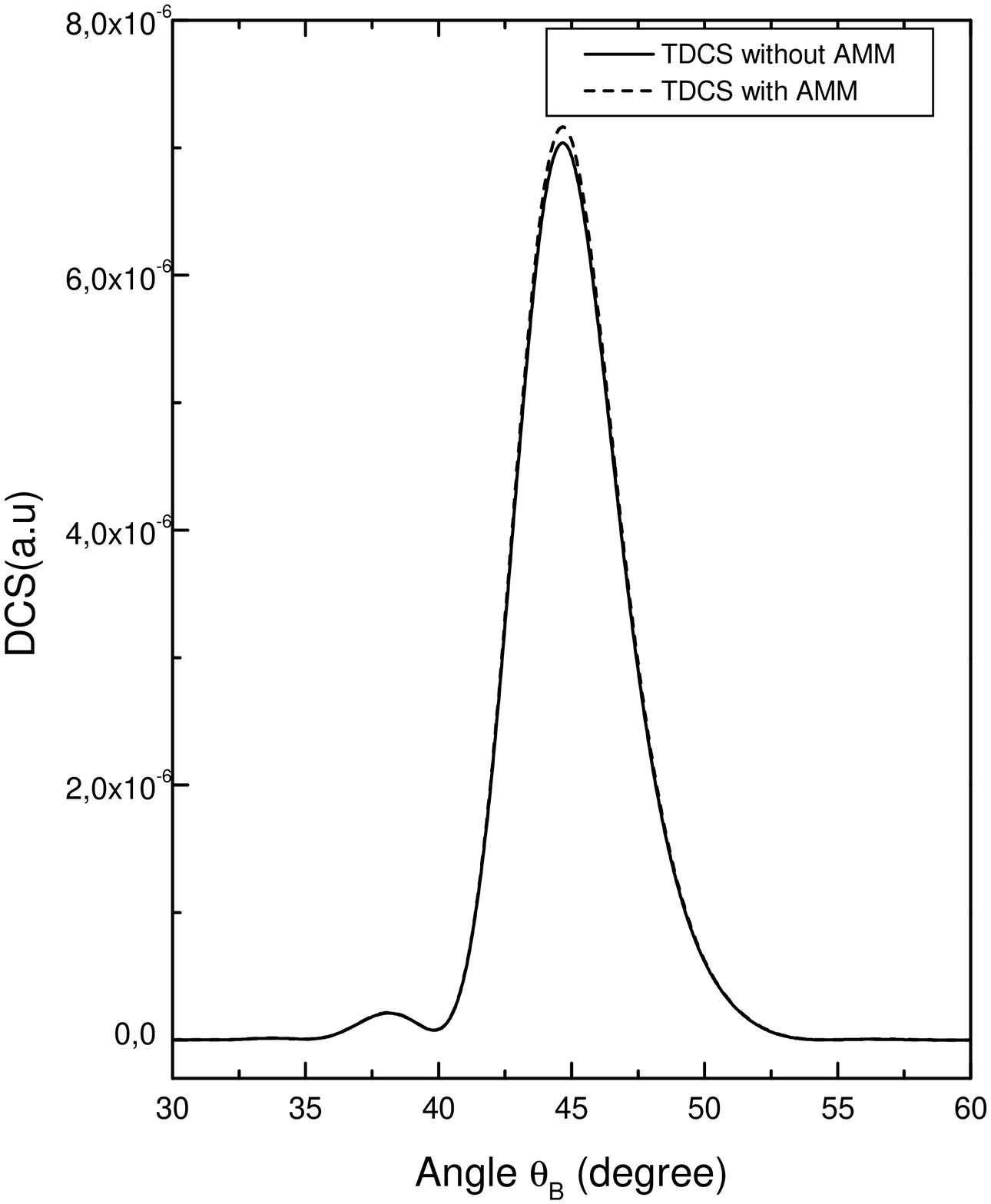}
\caption{The TDCSs as a function
of the angle $\theta_B$ for $s=1$ and $l_B=-1$ (we obtain the same figure
for $s=-1$ and $l_B=1$). The incident electron kinetic
energy is $T_i=2700 \;eV$ and the ejected electron kinetic energy is
 $T_B=1349.5 \;eV$. The geometric parameters are $\theta_i=0^{\circ}$, $\phi_i=\phi_f=0^{\circ}$, $\theta_f=45^{\circ}$ and $\phi_B=180^{\circ}$.  }
\end{figure}
\section{Results and discussions}
In this section, the results of the applications of the foregoing equations are presented by numerically evaluating the
TDCSs for the value of the electron's anomaly $\kappa = 1159652188.4\; 10^{-12}$ \cite{19}. We have chosen the angular frequency $\omega= 0.043\; a.u$ of a Nd:YAG laser. We have also discussed the laser-assisted TDCSs under three kinds of conditions : (a) without taking into account the AMM effect of all electrons (incident, scattered and ejected ), (b) taking into account only the AMM effect for the incident and scattered electrons, (c) finally, taking into account the AMM effects for all electrons. We choose a geometry where $p_i$ is along the $Oz$ axis ($\theta_i=\phi_i=0^{\circ}$). For the scattered electron,
($\theta_f=45^{\circ}$, $\phi_f=0^{\circ}$) and for the ejected electron $\phi_B=180^{\circ}$ and the angle $\theta_B$ varies approximatively from $30^{\circ}$ to $60^{\circ}$. This is an angular situation where we have a coplanar geometry.
\begin{figure}[!h]
\centering
\includegraphics[angle=0,width=3in,height=2.5 in]{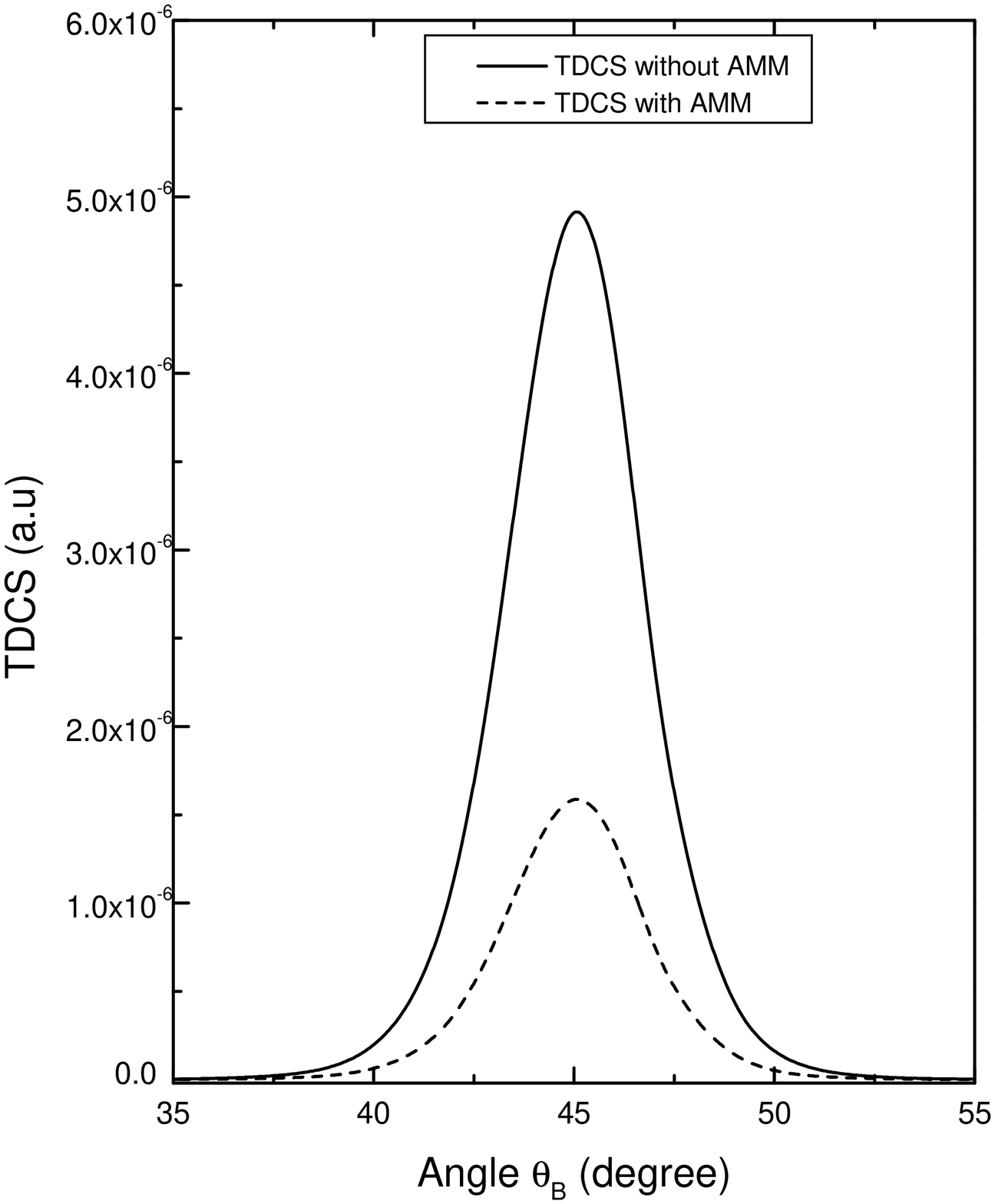}
\caption{The TDCSs as a function
of the angle $\theta_B$. The incident electron kinetic
energy is $T_i=5109 \;eV$ and the ejected electron kinetic energy is
 $T_B=2554.5 \;eV$. The electrical strength field is $\mathcal{E}=0.2\;a.u$ and the number of photons exchanged are $s=\pm 10$ and $l_B=\pm 10$}
\end{figure}
In Figure 1, we give the relation between the TDCSs and the angle of the ejected electron corresponding to three cases; the solid-line : results obtained by neglecting  the AMM effects of all electrons in the formalism, the long dash-line : results obtained by considering the AMM effects in the formalism but with the electron's anomaly $\kappa=0$ and the electrical field strength $\mathcal{E}=0\; a.u$. The dash-line : results obtained by using the plane waves. The results show that the three approaches give identical curves. In figure 2, we show the TDCS with and without AMM effects  for $s=1$ and $l_B=-1$. We have obtained the same curve for the case $s=-1$ and $l_B=1$. Once again this figure justifies clearly the accuracy and the consistency of our new formalism even if it contains a very long analytical formula which is not prone to calculation by hand.
\begin{figure}[!h]
\centering
\includegraphics[angle=0,width=3in,height=2.5 in]{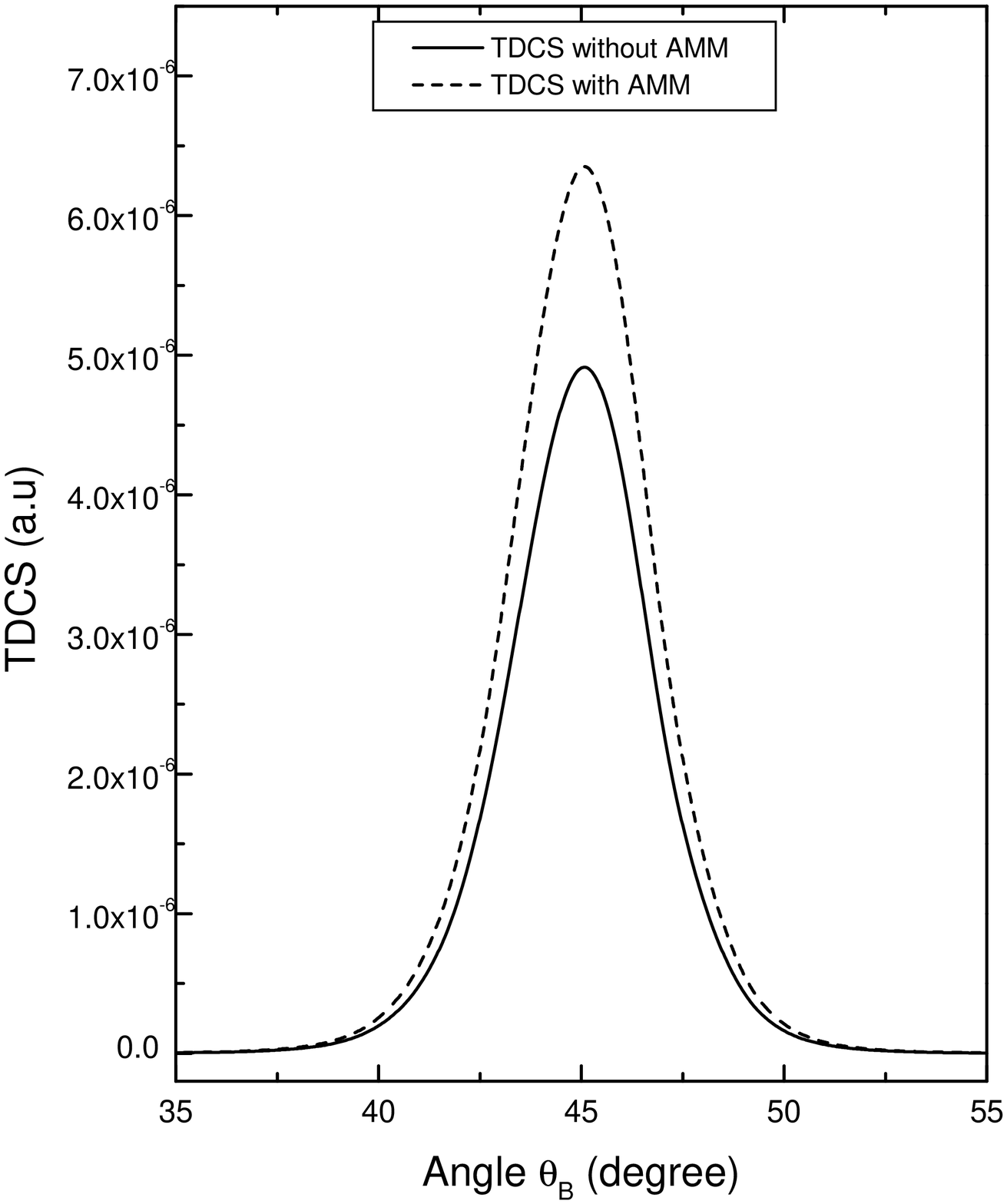}
\caption{The TDCSs as a function
of the angle $\theta_B$. The incident electron kinetic
energy is $T_i=5109 \;eV$ and the ejected electron kinetic energy is
 $T_B=2554.5 \;eV$. The electrical strength field is $\mathcal{E}=0.2\;a.u$ and the number of photons exchanged are $s=\pm 10$ and $l_B=\pm 10$ }
\end{figure}
Figure 3 illustrates the variation of TDCSs with AMM effects of the incident and scattered electrons versus the angle of the ejected electron.
\begin{figure}[!t]
\centering
\includegraphics[angle=0,width=4in,height=2.5 in]{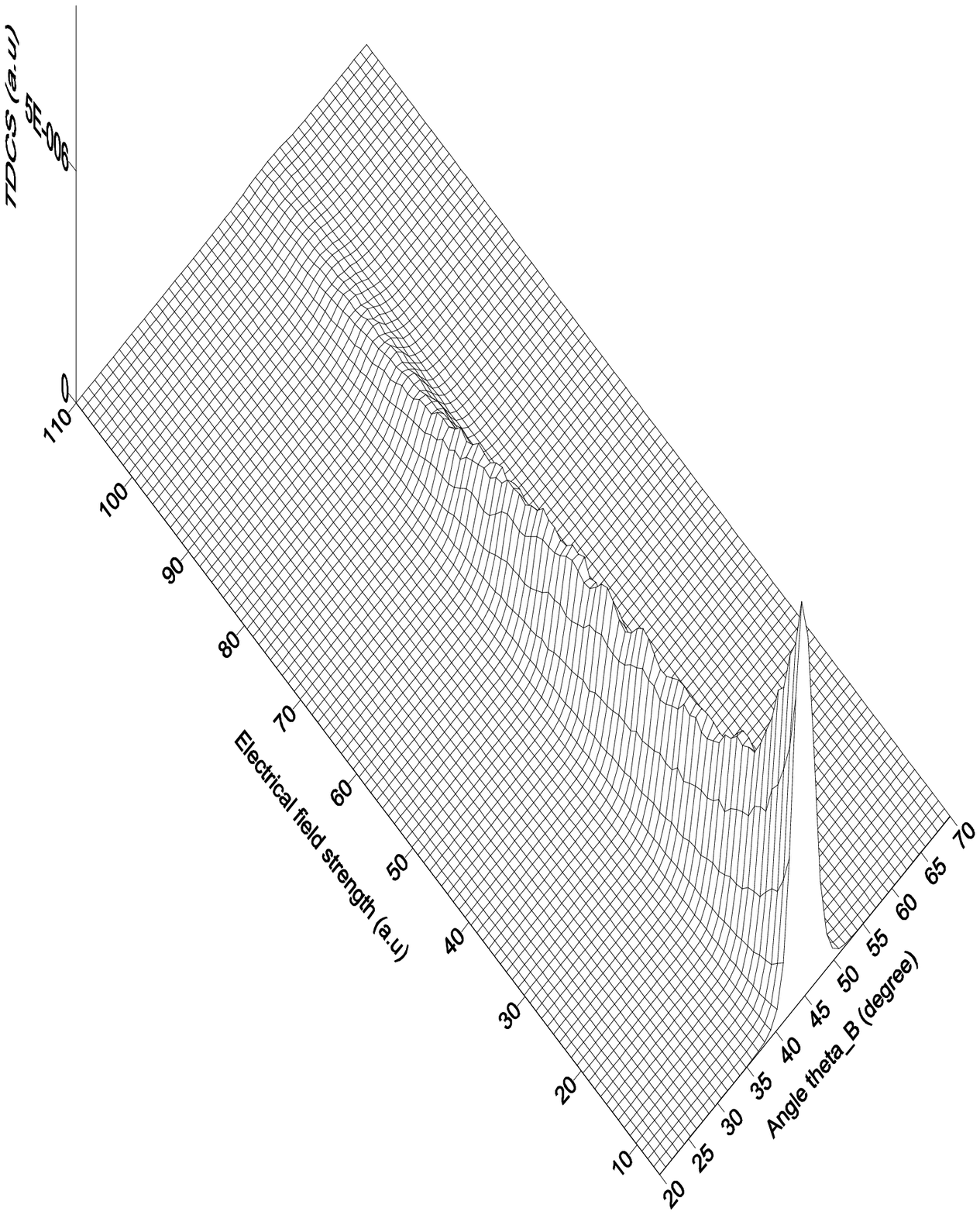}
\caption{The TDCS with AMM  as a function both
of the angle $\theta_B$ and the electrical field strength ($\mathcal{E}$ scaled in $10^{-2}$). The incident electron kinetic
energy is $T_i=5109 \;eV$ and the ejected electron kinetic energy is $T_B=2554.5 \;eV$. The geometric parameters are the same and the number of photons exchanged are $s=\pm 5$ and $l_B=\pm 5$ }
\end{figure}
 Three times magnitude between the two approaches is recognized. In our previous paper published earlier \cite{20}, the TDCS with electron's AMM effects always overestimates the TDCS without electron's AMM effects. This result is not justified if one considers only  the AMM effects of the incident and scattered electrons. Figure 4 shows the same dependence of the TDCS with AMM effects of all electrons (incident, scattered and ejected) and an emergent picture completely different . Indeed, the value of the TDCS with AMM effects at its maximum overestimates the TDCS without AMM. This means that, by introducing the AMM effects of the ejected electron, we have obtained a qualitative result similar to that obtained in our previous paper \cite{20}. For all energies, even if the process is very different $(e,2e)$, we have reached the same conclusion in which the TDCS with AMM always overestimates the TDCS without AMM.
Figure 5 shows a three-dimensional plot of the calculated triple differential cross section with the electron's anomalous magnetic moment effects. Two characteristic features of this landscape are obtained : first, the abrupt fall in the triple differential cross section at small and large angles; second, for the angles,  particular in the vicinity of ($\theta_B=45^{\circ}$) which represents the binary coplanar geometry,  the ejected electron loses its Coulombian behavior and the TDCS decreases with the intensity. We would like also to mention that the plane wave results should not be too reliable for the slower electrons, in this case the long-range Coulomb interaction should in no way be neglected. We are not in a position to compare our semirelativistic results with the existing theoretical works since the present theory is particularly meant for binary coplanar geometry and takes into account the electron's AMM effects whereas the other theoretical results refer to non relativistic case. Thus for a proper comparison we have to await the experimental data.
\section{Conclusion.}
In this paper, we have extended our treatment of the ionization of atomic hydrogen
by electronic impact in the presence of a circularly polarized laser field to the case
of the ionization with the introduction of the electron's anomalous magnetic moment effects.
The calculations have been performed in the framework of the first Born approximation
and in the binary coplanar geometry. These results show, notably
in comparison to more simplified approaches (TDCS without electron's AMM effects), the importance
of the full Dirac approach, especially in the case of
 intense laser fields and high energies. Important differences have been found
when the formalism of the triple differential cross sections with and without AMM is used.

\end{document}